\documentstyle[aps,prd,tighten,epsf]{revtex}
\begin{document}
\draft
\newcommand{\be}{\begin{equation}}
\newcommand{\ee}{\end{equation}}
\newcommand{\ben}{\begin{eqnarray}}
\newcommand{\een}{\end{eqnarray}}

\newcommand{\la}{{\lambda}}
\newcommand{\Om}{{\Omega}}
\newcommand{\ta}{{\tilde a}}
\newcommand{\bdel}{{\bar \delta}}
\newcommand{\bh}{{\bar h}}
\newcommand{\si}{{\sigma}}
\newcommand{\th}{{\theta}}
\newcommand{\C}{{\cal C}}
\newcommand{\D}{{\cal D}}
\newcommand{\cA}{{\cal A}}
\newcommand{\cT}{{\cal T}}
\newcommand{\cO}{{\cal O}}
\newcommand{\eeo}{\cO ({1 \over E})}
\newcommand{\G}{{\cal G}}
\newcommand{\cL}{{\cal L}}
\newcommand{\cH}{{\cal H}}
\newcommand{\cE}{{\cal E}}
\newcommand{\M}{{\cal M}}

\newcommand{\p}{\partial}
\newcommand{\na}{\nabla}
\newcommand{\ssum}{\sum\limits_{i = 1}^3}
\newcommand{\dssum}{\sum\limits_{i = 1}^2}
\newcommand{\tal}{{\tilde \alpha}}

\newcommand{\tp}{{\tilde \phi}}
\newcommand{\tPhi}{\tilde \Phi}
\newcommand{\tpsi}{\tilde \psi}
\newcommand{\tim}{{\tilde \mu}}
\newcommand{\tr}{{\tilde \rho}}
\newcommand{\tir}{{\tilde r}}
\newcommand{\rp}{r_{+}}
\newcommand{\hr}{{\hat r}}
\newcommand{\rv}{{r_{v}}}
\newcommand{\dr}{{d \over d \hr}}
\newcommand{\dR}{{d \over d R}}

\newcommand{\hhf}{{\hat \phi}}
\newcommand{\hhM}{{\hat M}}
\newcommand{\hhQ}{{\hat Q}}
\newcommand{\hht}{{\hat t}}
\newcommand{\hhr}{{\hat r}}
\newcommand{\hhS}{{\hat \Sigma}}
\newcommand{\hhD}{{\hat \Delta}}
\newcommand{\hhm}{{\hat \mu}}
\newcommand{\hro}{{\hat \rho}}
\newcommand{\hhz}{{\hat z}}

\newcommand{\tD}{{\tilde D}}
\newcommand{\tB}{{\tilde B}}
\newcommand{\tV}{{\tilde V}}
\newcommand{\hT}{\hat T}
\newcommand{\tF}{\tilde F}
\newcommand{\tT}{\tilde T}
\newcommand{\hC}{\hat C}
\newcommand{\ep}{\epsilon}
\newcommand{\bep}{\bar \epsilon}
\newcommand{\ppp}{\varphi}
\newcommand{\Ga}{\Gamma}
\newcommand{\ga}{\gamma}
\newcommand{\hth}{\hat \theta}
\title{First Law of Black Rings Thermodynamics in Higher Dimensional
Dilaton Gravity with $p + 1$ Strength Forms}

\author{Marek Rogatko}
\address{Institute of Physics \protect \\
Maria Curie-Sklodowska University \protect \\
20-031 Lublin, pl.Marii Curie-Sklodowskiej 1, Poland \protect \\
rogat@tytan.umcs.lublin.pl \protect \\
rogat@kft.umcs.lublin.pl}
\date{\today}
\maketitle
\smallskip
\pacs{ 04.50.+h, 98.80.Cq.}
\bigskip
\begin{abstract}
We derive the first law of black rings thermodynamics 
in $n$-dimensional Einstein dilaton gravity with additional $(p + 1)$-form field 
strength being the simplest generalization of five-dimensional theory
containing a stationary black ring solution with dipole charge. It was done by
means of choosing any cross section of the event
horizon to the future of the bifurcation surface. 
\end{abstract}
\baselineskip=18pt
\par
\section{Introduction}
Recently there has been a great resurgence of interests in $n$-dimen\-sio\-nal generalization
of black holes motivated by various attempts of constructing the unified theories.
As far as the static $n$-dimensional black holes is concerned the uniqueness theorem for them is
quite well established \cite{uniq}. But for stationary axisymmetric ones the problem is
much more complicated. In Ref.\cite{emp02} it was shown that even in five-dimensional
spacetime there is the so-called {\it black ring} solution possessing $S^2 \times S^1$
topology of the event horizon and having the same mass and angular momentum as spherical 
five-dimensional stationary axisymmetric black hole. 
On the contrary, when one assumes the spherical topology of the horizon $S^3$ the uniqueness
proof can be conducted (see Ref.\cite{mor04} for the vacuum case and Ref.\cite{rog04a} for the 
stationary axisymmetric
self-gravitating $\sigma$-model). As was remarked in \cite{hor05} due to {\it the vast number of
the ways of forming of these objects, one can take the view that the spirit of no-hair theorem
is preserved}.
\par
The black ring solution was vastly treated in literature, i.e., the solution possessing
the electric and magnetic dipole charge was found \cite{elv03,elv03b},
static black ring solution in five-dimensional Einstein-Maxwell-dilaton gravity 
was presented in \cite{kun05} and systematically derived in \cite{yaz05} both in asymptotically
flat and non-asymptotically flat case. Recently, the rotating non-asymptotically flat black ring
solution in charged dilaton gravity was given \cite{yaza}. Also the supersymmetric black ring 
solutions were discovered \cite{sup}.
\par
As far as the black hole thermodynamics is concerned, Wald \cite{wal94} referred two versions 
of the black hole thermodynamics. The first one, the so-called {\it
equilibrium state version} was treated for the first time
in the seminal paper of
Bardeen, Carter and Hawking \cite{bar73}. In this attitude the linear
perturbations of a stationary electrovac black hole to another stationary black hole
were taken into account. 
In Ref.\cite{sud92} arbitrary asymptotically flat perturbations of a stationary 
black hole were considered. The first law of black hole thermodynamics valid for
an arbitrary diffeomorphism invariant Lagrangian 
with metric and matter fields possessing stationary and axisymmetric
black hole solutions was given in Refs.\cite{wal93}-\cite{iye97}, while
the higher curvature terms and higher derivative terms in the metric
were considered in
\cite{jac}. The case when the Lagrangian is an arbitrary function of metric,
Ricci tensor and a scalar field was elaborated in Ref.\cite{kog98}.
Then, the case of a charged and rotating black hole
where fields were not smooth through the event horizon was treated in \cite{gao03}.
\par
On the other hand, the {\it physical process} version of the first law of black hole thermodynamics
was realized
by changing a stationary black hole by some infinitesimal physical process, e.g.,
when matter was thrown into black hole was considered. Assuming that the black hole eventually settles down to a 
stationary state we are able to 
find the changes of black hole's parameters and establish the first law of black hole mechanics.
The {\it physical process } version of the first 
law of black hole thermodynamics in Einstein theory was proved in Ref.\cite{wal94} and
generalized for Einstein-Maxwell (EM) black holes in Ref.\cite{gao01}. For 
Einstein-Maxwell axion-dilaton (EMAD) gravity black holes it was derived in Ref.
\cite{rog02}.\\
The first law of black hole thermodynamics was also intensively studied in the case of $n$-dimensional
black holes. The {\it equilibrium state} version was elaborated in Ref.\cite{equi1} 
under the assumption of spherical topology of black holes. Some of the works assume that four-dimensional black 
hole uniqueness theorem extends to higher dimensional case \cite{equi2}.
The {\it physical process} of the first law of black hole thermodynamics in $n$-dimension was 
treated in Ref.\cite{rog05}.\\
As far as the black ring first law of mechanics is concerned, 
the general form of this law was established in Ref.\cite{cop05}, using the notion of
bifurcate Killing horizons and taking into account dipole charges.
The
{\it physical process} version of the first law of thermodynamics
in the higher dimensional gravity containing $(p + 1)$-form field strength and dilaton fields which
constitutes the simplest generalization of five-dimensional one, which in turn contains stationary 
black ring solution with dipole charge \cite{elv05}, was given in Ref.\cite{rog05br}.
\par
In our paper we shall derive the first law of black ring mechanics choosing an arbitrary cross
section of the event horizon to the future of the bifurcation surface, contrary to the previous
derivations based on taking into account the bifurcation surface as the boundary of 
the hypersurface extending to spatial infinity. This attitude enables one to treat fields which 
are not necessarily smooth through the horizon \cite{gao03}, i.e., one requires only
that the pullback of the fields under considerations in the future of the bifurcation surface
be smooth.

\section{The first law of black ring mechanics}
We shall consider the simplest higher dimensional generalization of
of the five-dimensional theory with three-form field that admits stationary
black ring solutions. Namely, it will be subject to the relation 
\be
{\bf L } = {\bf \ep} \bigg(
{}^{(n)}R - {1 \over 2} \na_{\mu}\phi \na^{\mu} \phi - {1\over 2 (p + 1)!} e^{-{\alpha} \phi}
H_{\mu_{1} \dots \mu_{p+1}} H^{\mu_{1} \dots \mu_{p+1}}
\bigg),
\label{lag}
\ee
where $ {\bf \ep}$ is the $n$-dimensional volume element,
$\phi$ constitutes the dilaton field while
$H_{\mu_{1} \dots \mu_{p+1}} = (p + 1)! \na_{[ \mu_{1}} B_{{\mu_{2} \dots \mu_{p+1}]}} $ is $(p + 1)$-form field strength.
\par
The symplectic $(n - 1)$-form
$\Theta_{j_{1} \dots j_{n-1}}[\psi_{\alpha}, \delta \psi_{\alpha}]$, was calculated in \cite{rog05br}.
It yields
\be
\Theta_{j_{1} \dots j_{n-1}}[\psi_{\alpha}, \delta \psi_{\alpha}] =
\ep_{\mu j_{1} \dots j_{n-1}} \bigg[
\omega^{\mu} - e^{-{\alpha} \phi} H^{\mu \nu_{2} \dots \nu_{p+1}}~\delta B_{\nu_{2} \dots \nu_{p+1}}
- \na^{\mu} \phi~ \delta \phi \bigg],
\ee 
where $\omega_{\mu} = \na^{\alpha} \delta g_{\alpha \mu} - \na_{\mu} 
\delta g_{\beta}{}{}^{\beta}$. For brevity,
we denote fields in the underlying theory by $\psi_{\alpha}$,
while their variations by $\delta \psi_{\alpha}$.
The Noether charges are as follows:
\be
Q_{j_{1} \dots j_{n-2}}^{GR}(\xi) = - \ep_{j_{1} \dots j_{n-2} a b} \na^{a} \xi^{b},
\ee
while $Q_{j_{1} \dots j_{n-2}}^{B}$ has the following form:
\be
Q_{j_{1} \dots j_{n-2}}^{B}(\xi) = {p \over (p + 1)!} \ep_{m \alpha j_{1} \dots j_{n-1}}
~\xi^{d}~B_{d \alpha_{3} \dots \alpha_{p+1}}~ e^{-{\alpha} \phi} H^{m \alpha \alpha_{3} \dots \alpha_{p+1}}.
\ee
Let us consider the case when the spacetime satisfies asymptotic conditions at infinity
and the Killing vector $\xi^{\mu}$ 
guarantees an asymptotic symmetry. Then, there exists a conserved quantity $H_{\xi}$ associated
with the Killing vector fields under consideration \cite{wal00}, which implies
\be
\delta H_{\xi} = \int_{\infty} \bigg( \bdel Q(\xi) - \xi \Theta \bigg),
\label{qua}
\ee
where $\bdel$ denotes the variation which has no effect on $\xi_{\alpha}$ since the Killing
vector field is treated as a fixed background and it ought not to be varied in expression (\ref{qua}).\\ 
For the case when a hypersurface $\Sigma$ extends to infinity and has an inner boundary
$\p \Sigma$, and moreover $\xi_{\alpha}$ is a symmetry of all dynamical fields as well as $\psi_{\alpha}$
and $\delta \psi_{\alpha}$ fulfill the linearized equations of motion, then it follows
\cite{iye94} that the integral over infinity can be changed into the inner boundary one.
In our consideration we shall confine our attention to stationary and axisymmetric black ring
so the Killing vector field is of the form
\be
\xi^{\mu} = t^{\mu} + \sum_{i} \Omega_{(i)} \phi^{\mu (i)},
\ee
where $\phi^{\mu (i)}$ are the Killing vectors responsible for the rotation in the adequate 
directions. As we take $\xi^{\alpha}$ to be an asymptotical time translation
$t^{\alpha}$, we obtain from Eq.(\ref{qua}) the variation of canonical energy
$\delta {\cal E}$
and taking into account $\phi^{\mu (i)}$ we get variations of the adequate canonical angular 
momentum $\delta {\cal J}_{(i)}$.
\par
From this stage on, we shall take into account an asymptotically hypersurface $\Sigma$
terminating on the portion of the event horizon ${\cal H}$ to the future of the bifurcation
surface. By $S_{\cal H}$ we denote the cross section of the event horizon which
constitutes the inner boundary of the hypersurface $\Sigma$. We take into account a variation
between two neighbouring states of black rings.
When one compares two slightly different solutions there is a freedom in which points
they can be chosen to correspond. We choose this freedom to make $S_{\cal H}$ the same
of the two solutions (freedom of the general coordinate transformation) as well as the 
null vector remains normal to it. The
stationarity and 
axisymmetricity described by the Killing vector fields 
$t^{\alpha}$ and $\phi^{\alpha (i)}$ will also be conserved. Thus, the perturbations 
of the Killing vector fields $\delta t^{\alpha}$ and $\delta \phi^{\alpha (i)}$ 
will be equal to zero, which implies in turn that the
corotating Killing vector field is assigned to the expression
\be
\delta \xi^{\mu} =  \sum_{i} \delta \Omega_{(i)} \phi^{\mu (i)}.
\ee
Let us assume that $(g_{\mu \nu}, B_{\alpha_{1} \dots \alpha_{p}}, \phi)$ are solutions 
of the equations of motion derived from the Lagrangian (\ref{lag}) and 
$(\delta g_{\mu \nu},~\delta B^{\alpha_{1} \dots \alpha_{p}},~\delta \phi)$
are the linearized perturbations satisfying Eqs. of motion. We require that the pullback
of $B_{\alpha_{1} \dots \alpha_{p}}$ to the future of the bifurcation surface be smooth, but not 
necessarily smooth on it \cite{gao03}. One assumes further, that $B_{\alpha_{1} \dots \alpha_{p}}$
and $\delta B_{\alpha_{1} \dots \alpha_{p}}$ fall off sufficiently rapid at infinity.
Then, those fields do not contribute to the canonical energy and canonical momenta.
It implies
\be
\alpha~ \delta M -  \sum_{i} \Omega_{(i)} \delta J^{(i)} = 
\int_{S_{\cal H}} \bigg( \bdel Q(\xi) - \xi \Theta \bigg),
\ee
where $\alpha = {n-3 \over n-2}$.\\ 
First we calculate the integral over the symplectic $(n-1)$-form bounded with the dilaton field.
We express $\ep_{\mu a j_{1} \dots j_{n-2}}$ by the volume element on $S_{\cal H}$ and
by the vector $N^{\alpha}$, the {\it ingoing} future directed null normal to $S_{\cal H}$, which is normalized
to $N^{\alpha} \xi_{\alpha} = - 1$ \cite{walbook}.
It gives the expression written as
\ben
\int_{S_{\cal H}}\xi^{j_{1}}~ \Theta_{j_{1} \dots j_{n-1}}^{\phi} =
\int_{S_{\cal H}} \ep_{j_{1} \dots j_{n-2}} N_{\alpha} \xi^{\alpha}~ \xi_{\mu} \na^{\mu} \phi~
\delta \phi = 0,
\een
where we used the fact that $\cL_{\xi} \phi = 0$.\\
Consequently, we turn now our attention to the $(p + 1)$-form field.
In order to find the integral over the event horizon we take into account the following relation:
\ben \label{cf}
p!~\cL_{\xi} B_{\alpha_{2} \dots \alpha_{p+1}}~H^{m \alpha_{2} \dots \alpha_{p+1}}
&-& \xi^{d}~H_{d \alpha_{2} \dots \alpha_{p+1}}~H^{m \alpha_{2} \dots \alpha_{p+1}} \\ \nonumber
&=& p~ p! \na_{\alpha_{2}} \bigg(
\xi^{d}~B_{d \alpha_{3} \dots \alpha_{p+1}}
\bigg)~H^{m \alpha_{2} \dots \alpha_{p+1}}.
\een
The first term of the left-hand side of Eq.(\ref{cf}) is equal to zero since 
the Killing vector field $\xi_{\alpha}$ constitutes the 
symmetry of the background solution. Next, let us consider $n$-dimensional Raychauduri 
equation written as follows:
\be
{d \theta \over d \lambda} = - {\theta^{2} \over (n - 2)} - \sigma_{ij} \sigma^{ij}
- R_{\mu \nu} \xi^{\mu} \xi^{\nu},
\label{ray}
\ee
where $\lambda$ denotes the affine parameter corresponding to vector $\xi_{\alpha}$, $\theta$ is the expansion and
$\sigma_{ij}$ is shear. Shear and expansion vanish in the stationary background, so 
the inspection of Eq.(\ref{ray}) yields that
$R_{\alpha \beta} \xi^{\alpha} \xi^{\beta} \mid_{\cH} = 0$, implying the following:
\be
{1 \over 2}\xi^{\mu} \na_{\mu} \phi~ \xi^{\nu} \na_{\nu} \phi +
{1 \over 2 p!} e^{- \alpha \phi}
H_{\mu \mu_{2} \dots \mu_{p+1}} H_{\nu}{}{}^{ \mu_{2} \dots \mu_{p+1}} \xi^{\mu} \xi^{\nu} \mid_{\cH} = 0.
\ee
Due to the fact that $\cL_{\xi} \phi = 0$,
it is easily seen that,
$H_{\mu}{}{}^{ \mu_{2} \dots \mu_{p+1}} \xi^{\mu} = 0$. Because of 
$H_{\mu \mu_{2} \dots \mu_{p+1}} \xi^{\mu} \xi^{\mu_{2}} = 0$, 
and having in mind
asymmetry of $H_{ \mu_{1} \dots \mu_{p+1}}$
one draws a conclusion that
$H_{\mu \mu_{2} \dots \mu_{p+1}} \xi^{\mu} \sim \xi_{\mu_{2}} \dots \xi_{\mu_{p+1}}$.
Just the pullback of $H_{\mu}{}{}^{ \mu_{2} \dots \mu_{p+1}} \xi^{\mu}$ to the event horizon is equal to zero,
which in turn constitutes that
$\xi^{d}~B_{d \alpha_{2} \dots \alpha_{p+1}}$ is a closed
$p$-form on the event horizon. Applying the Hodge theorem (see e.g., \cite{wes81}) it may be rewritten
as a sum of an exact and harmonic form. An exact form does not contribute to the above expression
because of the Eqs. of motion are satisfied. 
The harmonic part of
$\xi^{d}~B_{d \alpha_{2} \dots \alpha_{p+1}}$
gives the only contribution. Just, having in mind the duality between homology
and cohomology, one can conclude that
there is a harmonic form $\eta$ dual to $n - p- 1$ cycle $\cal S$ in the sense of the equality
of the adequate surface integrals. Then, it follows that the surface term will be of the form as
$\Phi_{l}~q_{l}$, where $\Phi_{l}$ is the constant relating to the harmonic part
of $\xi^{d}~B_{d \alpha_{2} \dots \alpha_{p+1}}$ and $q_{l}$ is a local charge \cite{cop05}.
These considerations allow one to write down the following:
\be
\int_{S_{\cal H}} Q_{j_{1} \dots j_{n-2}}^{B}(\xi) = \Phi_{l}~q_{l}.
\ee
Let us compute the variation $\bdel$. It reduces to
\ben
\bdel \int_{S_{\cal H}} Q_{j_{1} \dots j_{n-2}}^{B} (\xi) &=&
\delta \int_{S_{\cal H}} Q_{j_{1} \dots j_{n-2}}^{B} (\xi) -
\int_{S_{\cal H}} Q_{j_{1} \dots j_{n-2}}^{B} (\delta \xi) = 
\delta \bigg( \Phi_{l}~q_{l} \bigg) \\ \nonumber 
&-&
{p \over (p + 1)!}\int_{S_{\cal H}} \sum_{i} \delta \Omega_{(i)} \phi^{\mu (i)}
B_{\mu \alpha_{3} \dots \alpha_{p+1}} 
\ep_{m \alpha j_{1} \dots j_{n-2}}~ e^{-{\alpha} \phi}~ 
H^{m \alpha \alpha_{3} \dots \alpha_{p+1}}. 
\een
As immediate consequences of the above expressions one has
\ben \label{pt}
 \delta \Phi_{l}~q_{l} =
{p \over (p + 1)!}\int_{S_{\cal H}} \sum_{i} \delta \Omega_{(i)} \phi^{\mu (i)}
B_{\mu \alpha_{3} \dots \alpha_{p+1}} 
\ep_{m \alpha j_{1} \dots j_{n-2}}~ e^{-{\alpha} \phi}~ 
H^{m \alpha \alpha_{3} \dots \alpha_{p+1}} \\ \nonumber
+ {p \over (p + 1)!}\int_{S_{\cal H}} \xi^{d}~\delta B_{d \alpha_{3} \dots \alpha_{p+1}}
~N_{m}~\xi_{\alpha}~e^{-{\alpha} \phi}~ 
H^{m \alpha \alpha_{3} \dots \alpha_{p+1}}. 
\een
Further, we take into account symplectic $(n-1)$-form for the potential
$B_{\nu_{1} \dots \nu_{p}}$
\be
\int_{S_{\cal H}} \xi^{j_{1}}~\Theta_{j_{1} \dots j_{n-1}}^{B} =
- \int_{S_{\cal H}} \ep_{\mu j_{1} \dots j_{n-1}} 
e^{-{\alpha} \phi} H^{\mu \nu_{2} \dots \nu_{p+1}}~\delta B_{\nu_{2} \dots \nu_{p+1}}.
\ee
Using the fact that on the event horizon of black ring 
$H_{\mu \mu_{2} \dots \mu_{p+1}} \xi^{\mu} \sim \xi_{\mu_{2}} \dots \xi_{\mu_{p+1}}$
and expressing $\ep_{\mu a j_{1} \dots j_{n-2}}$ in the same form
as in the above case, one arrives at the following:
\be
\int_{S_{\cal H}} \xi^{j_{1}}~\Theta_{j_{1} \dots j_{n-1}}^{B} =
{p \over (p+1)!}
\int_{S_{\cal H}} \ep_{j_{1} \dots j_{n-2}}~ e^{-{\alpha} \phi}~ 
\xi_{\alpha}~ H^{\delta \alpha \nu_{3} \dots \nu_{p+1}}~N_{\delta}~
\xi^{\nu_{2}}~ \delta B_{\nu_{2} \dots \nu_{p+1}}.  
\label{bb1}
\ee
Combining Eq.(\ref{pt}) with the expressions (\ref{bb1}) 
we finally conclude
\be
\bdel \int_{S_{\cal H}} Q_{j_{1} \dots j_{n-2}}^{B} (\xi)
- \xi^{j_{1}}~\Theta_{j_{1} \dots j_{n-1}}^{B} = 
\Phi_{l}~\delta q_{l}.
\label{char}
\ee
Let us turn our attention to 
the contribution connected with gravitational
field. Namely, we begin with $Q_{j_{1} \dots j_{n-2}}^{GR} (\xi)$
and following the standard procedure \cite{bar73}
one arrives at the following:
\be
\int_{S_{\cal H}} Q_{j_{1} \dots j_{n-2}}^{GR} (\xi) = 2 \kappa A,
\ee
where $A = \int_{S_{\cal H}} \ep_{j_{1} \dots j_{n-2}}$ 
is the area of the black ring horizon. It now follows that  
\be
\bdel \int_{S_{\cal H}} Q_{j_{1} \dots j_{n-2}}^{GR} (\xi) =
2 \delta \bigg( \kappa A \bigg) + 2 \sum_{i} \delta \Omega_{(i)}~ J^{(i)},
\ee
where $J^{(i)}= {1 \over 2}\int_{S_{\cal H}}\ep_{j_{1} \dots j_{n-2} a b} \na^{a} \phi^{(i)b}$ 
is the angular momentum connected with the Killing vector
field $\phi_{(i)}$ responsible for the rotation in the adequate directions.
Having in mind calculations conducted in Ref.\cite{bar73} it could be verified that
the following integral is satisfied:
\be
\int_{S_{\cal H}} \xi^{j_{1}}~ Q_{j_{1} \dots j_{n-2}}^{GR} (\xi) =
2 A~ \delta \kappa + 2 \sum_{i} \delta \Omega_{(i)}~ J^{(i)}.
\ee
The above yields the conclusion that
\be
\bdel \int_{S_{\cal H}} Q_{j_{1} \dots j_{n-2}}^{GR} (\xi)
- \xi^{j_{1}}~\Theta_{j_{1} \dots j_{n-1}}^{GR} = 
2 \kappa~\delta A.
\label{arr}
\ee
On using Eqs.(\ref{char}) and (\ref{arr}),
we find that we have obtained the first law of black rings mechanics in Einstein
$n$-dimensional gravity with additional $(p+1)$-form field strength and dilaton
fields, in theory which is the simplest generalization of five-dimensional
one containing a stationary black ring solution with dipole charge. The first
law of black ring mechanics yields the following: 
\be
\alpha~ \delta M -  \sum_{i} \Omega_{(i)} \delta J^{(i)} 
+ \Phi_{l}~\delta q_{l} = 2 \kappa~\delta A.
\ee

\noindent
{\bf Acknowledgements:}\\
MR was supported in part by the Polish Ministry of Science and Information Society Technologies grant 1 P03B 049 29.



\begin{references}
%
\def\cmp#1#2#3{{ Commun. Math. Phys.} {\bf #1}, #2 (#3)}
\def\lmp#1#2#3{{ Lett. Math. Phys.} {\bf #1}, #2 (#3)}
\def\hpa#1#2#3{{ Hell. Phys. Acta} {\bf #1}, #2 (#3)}
\def\grg#1#2#3{{ Gen. Rel. Grav.} {\bf #1}, #2 (#3)}
\def\pr#1#2#3{{ Phys. Rev.} {\bf #1}, #2 (#3)}
\def\prl#1#2#3{{ Phys. Rev. Lett.} {\bf #1}, #2 (#3)}
\def\prd#1#2#3{{ Phys. Rev. D} {\bf #1}, #2 (#3)}
\def\pl#1#2#3{{ Phys. Lett} {\bf #1}, #2 (#3)}
\def\pla#1#2#3{{ Phys. Lett. A} {\bf #1}, #2 (#3)}
\def\plb#1#2#3{{ Phys. Lett. B} {\bf #1}, #2 (#3)}
\def\prep#1#2#3{{ Phys. Reports} {\bf #1}, #2 (#3)}
\def\phys#1#2#3{{ Physica} {\bf #1}, #2 (#3)}
\def\jcp#1#2#3{{ J. Comput. Phys.} {\bf #1}, #2 (#3)}
\def\jmp#1#2#3{{ J. Math. Phys.} {\bf #1}, #2 (#3)}
\def\jpm#1#2#3{{ J. Phys. A: Math. Gen.} {\bf #1}, #2 (#3)}
\def\cpr#1#2#3{{ Computer Phys. Rept.} {\bf #1}, #2 (#3)}
\def\cqg#1#2#3{{ Class. Quantum Grav.} {\bf #1}, #2 (#3)}
\def\cma#1#2#3{{ Computers Math. Applic.} {\bf #1}, #2 (#3)}
\def\mc#1#2#3{{ Math. Compt.} {\bf #1}, #2 (#3)}
\def\apj#1#2#3{{ Astrophys. J.} {\bf #1}, #2 (#3)}
\def\apjs#1#2#3{{ Astrophys. J. Suppl.} {\bf #1}, #2 (#3)}
\def\acta#1#2#3{{ Acta Astronomica} {\bf #1}, #2 (#3)}
\def\apl#1#2#3{{Ann. Physik. (Leipzig)} {\bf #1}, #2 (#3)}
\def\anp#1#2#3{{Ann. Phys. } {\bf #1}, #2 (#3)}
\def\sa#1#2#3{{ Sov. Astro.} {\bf #1}, #2 (#3)}
\def\sia#1#2#3{{ SIAM J. Sci. Statist. Comput.} {\bf #1}, #2 (#3)}
\def\aa#1#2#3{{ Astron. Astrophys.} {\bf #1}, #2 (#3)}
\def\mnras#1#2#3{{ Mon. Not. R. astr. Soc.} {\bf #1}, #2 (#3)}
\def\npb#1#2#3{{ Nucl. Phys. B} {\bf #1}, #2 (#3)}
\def\prsla#1#2#3{{ Proc. R. Soc. London, Ser. A} {\bf #1}, #2 (#3)}
\def\jhep#1#2#3{{ JHEP} {\bf #1}, #2 (#3)}
\def\nuc#1#2#3{{Nuovo Cimento B } {\bf #1}, #2 (#3)}
\def\ijmp#1#2#3{{Int. J. Mod. Phys. D} {\bf #1}, #2 (#3)}
\def\atmp#1#2#3{{Adv. Theor. Math. Phys.} {\bf #1}, #2 (#3)}
\def\ptps#1#2#3{{Prog. Theor. Phys. Suppl.} {\bf #1}, #2 (#3)}
\def\lmp#1#2#3{{Lett. Math. Phys. } {\bf #1}, #2 (#3)}
%
\def\hepph#1#2{{ hep-ph }{\bf #1} (#2)}
\def\hepth#1#2{{ hep-th }{\bf #1} (#2)}
\def\grqc#1#2{{ gr-qc }{\bf #1} (#2)}
\def\ibid#1#2#3{{ {\it ibid.} }{\bf #1}, #2 (#3)}
%

\bibitem{uniq}
G.W.Gibbons, D.Ida, and T.Shiromizu, \ptps{148}{284}{2003},\\
G.W.Gibbons, D.Ida, and T.Shiromizu, \prd{66}{044010}{2002},\\
G.W.Gibbons, D.Ida, and T.Shiromizu, \prl{89}{041101}{2002},\\
M.Rogatko, \cqg{19}{L151}{2002},\\
M.Rogatko, \prd{67}{084025}{2003},\\
M.Rogatko, \ibid{70}{044023}{2004}.
\bibitem{emp02}
R.Emparan and H.S.Reall, \prl{88}{101101}{2002}.
\bibitem{mor04}Y.Morisawa and D.Ida, \prd{69}{124005}{2004}.
\bibitem{rog04a}M.Rogatko, \prd{70}{084025}{2004}.
\bibitem{hor05}G.T.Horowitz, {\it Higher Dimensional Generalization of the Kerr Black Hole}
\grqc{0507080}{2005}.
\bibitem{elv03}
H.Elvang, \prd{68}{124016}{2003}.
\bibitem{elv03b}
H.Elvang and R.Emparan, \jhep{11}{035}{2003}.
\bibitem{kun05}
H.K.Kunduri and J.Lucietti, \plb{609}{143}{2005}.
\bibitem{yaz05}S.Yazadjiev, \cqg{22}{3875}{2005}.
\bibitem{yaza}S.Yazadjiev, \prd{72}{104014}{2005}. 
\bibitem{sup}
I.Bena and P.Kraus, \prd{70}{046003}{2004},\\
I.Bena, \prd{70}{105018}{2004},\\
H.Elvang, R.Emparan, D.Mateos, and H.S.Reall, \prl{93}{211302}{2004},\\
J.P.Gauntlett, J.B.Gutowski, C.M.Hull, S.Pakis, and H.S.Reall, \cqg{20}{4587}{2003},\\
J.P.Gauntlett and J.B.Gutowski, \prd{71}{045002}{2005}.
\bibitem{wal94}
R.M.Wald, {\it Quantum Field Theory in Curved Spacetime and Black Hole Thermodynamics}, 
University of Chicago Press (Chicago, 1994).
\bibitem{bar73}
J.M.Bardeen, B.Carter and S.W.Hawking, \cmp{31}{161}{1973}.
\bibitem{sud92}
D.Sudarsky and R.M.Wald, \prd{46}{1453}{1992}.
\bibitem{wal93}
R.M.Wald, \prd{48}{R3427}{1993}.
\bibitem{iye94}
V.Iyer and R.M.Wald, \prd{50}{846}{1994}.
\bibitem{iye95}
V.Iyer and R.M.Wald, \prd{52}{4430}{1995}.
\bibitem{iye97}
V.Iyer, \prd{55}{3411}{1997}.
\bibitem{jac}
T.Jacobson, G.Kang, and R.C.Myers, \prd{49}{6587}{1994},\\
T.Jacobson, G.Kang, and R.C.Myers, \prd{52}{3518}{1995}.
\bibitem{kog98}
J.Koga and K.Maeda, \prd{58}{064020}{1998}.
\bibitem{gao03} 
S.Gao, \prd{68}{044016}{2003}.
\bibitem{gao01}
S.Gao and R.M.Wald, \prd{64}{084020}{2001}.
\bibitem{rog02}
M.Rogatko, \cqg{19}{3821}{2002}.
\bibitem{equi1}
G.W.Gibbons, M.J.Perry, and C.N.Pope, \cqg{22}{1503}{2005},\\
M.Korzynski, J.Lewandowski, and T.Pawlowski, \ibid{22}{2001}{2005},\\
M.Rogatko, \prd{71}{024031}{2005}.
\bibitem{equi2}
R.C.Myers and M.J.Perry, \anp{172}{304}{1986},\\
J.P.Gauntlett, R.C.Myers, and P.K.Townsend, \cqg{16}{1}{1999},\\
P.K.Townsend and M.Zamaklar, \cqg{18}{5269}{2001}.
\bibitem{rog05}M.Rogatko, \prd{71}{104004}{2005}.
\bibitem{cop05}
K.Copsey and G.T.Horowitz, {\it The Role of Dipole Charges in Black Hole
Thermodynamics}, \hepth{0505278}{2005}.
\bibitem{elv05}
H.Elvang, R.Emparan, and P.Figueras, \jhep{02}{031}{2005}.
\bibitem{rog05br}M.Rogatko, \prd{72}{074008}{2005}, Erratum \ibid{72}{089901}{2005}.
\bibitem{wal00}R.M.Wald and A.Zoupas, \prd{61}{084027}{2000}.
\bibitem{walbook}R.M.Wald, {\it General Relativity} (University of Chicago Press,
Chicago, 1984).


\bibitem{wes81}
C.von Westenholz, {\it Differential Forms in Mathematical Physics}, North-Holland
Publishing (New York, 1981).

\end{references}
\end{document}